\documentclass[12pt]{article}
\usepackage{geometry}             
\usepackage{graphicx}
\usepackage{amssymb}
\usepackage{amsmath}
\usepackage{epstopdf}
\usepackage{comment}

\usepackage[hyperindex=true,
          pdfstartview=FitH,
          bookmarksnumbered=true,
          bookmarksopen=true,
          citecolor=blue,
          linkcolor=blue,
          colorlinks=true,
          unicode]{hyperref}

\begin{document}

\title{Strange Lagrangian systems and \\
statistical mechanics}
\author{Liu Zhao\\
School of Physics, Nankai University, Tianjin 300071, China\\
email: {\it lzhao@nankai.edu.cn}}
\date{}                             
\maketitle

\begin{abstract}
We consider the canonical ensemble of $N$ particles admitting a strange  
Hamiltonian description. Each of the particles obeys a set of Newtonian equation 
of motion, which can also be described by the standard canonical Hamiltonian 
mechanics. However, the thermodynamics corresponding to
the strange description and canonical description differ drastically from each 
other. In other words, the strange description and the standard
canonical description are inequivalent on the level of thermodynamics.
\end{abstract}

\section{Introduction}

Canonical ensemble in statistical mechanics is deeply rooted from 
Hamiltonian mechanics. The intimate relationship between statical mechanics and 
Hamiltonian mechanics is best illustrated in the definition of the 
canonical partition function 
\begin{align}
P=\mathcal{Z}^N=\left(\frac{1}{h}\int \mathrm{e}^{-\beta H} d\Omega\right)^N, 
\label{part}
\end{align}
where $\beta=1/kT$ as usual, $N$ is the number of degrees of freedom in the system, 
$H$ and $d\Omega$ are respectively the Hamiltonian and the symplectic volume 
element for a single degree of freedom in the system. It is evident that the 
partition function is invariant under the symplectic group of transformations
on the phase space of the underlying Hamiltonian system. 

Nowadays it is well known that one can often associate many different Hamiltonian 
descriptions to the same set of Hamiltonian equations of motion. Under some mild 
assumptions, it was argued in \cite{Ercolessi:2001p2304} that different Hamiltonian 
descriptions give rise to the same canonical partition function in statistical 
mechanics, yielding the same set of thermodynamical quantities. This novel 
invariance is beyond the standard symplectic invariance which keeps the 
Hamiltonian structure, so it seems worth to pay some extra attention toward 
it. In particular, is tempting to ask what happens when the assumptions used in 
\cite{Ercolessi:2001p2304} are violated. 

Please note that, for systems
possessing multiple Hamiltonian descriptions, there have been debates in the
literature on the proper choice of Hamiltonian functions. Some people insist that
the physical Hamiltonian must be decomposable into kinetic and potential energy 
parts. However, as far as the equations of motion are concerned, we don't 
seem to be able to justify which description is superior over the others. This 
indistinguishability between different Hamiltonian descriptions is uplifted
by the work of \cite{Ercolessi:2001p2304} to the level of statistical mechanics.

On the other hand, it has become clear only recently that,
for some special kind of mechanical systems, there exist choices of 
Hamiltonian structures in which certain fundamental aspects of classical canonical 
Hamiltonian mechanics are changed. In \cite{Shapere:2012p2114}, Shapere and Wilczek
found some Lagrangian systems for which the Hamiltonian become 
multivalued in terms of the canonical phase space variables. 
A direct consequence of this multivaluedness is that the time translation symmetry
is spontaneously broken in the ground states, even though the Hamiltonian itself 
is conserved. For an ensemble of 
such mechanical systems, the canonical partition function as given in (\ref{part})
is ill defined, because $H$ and $d\Omega$ are not well defined 
simultaneously. However, as explored in \cite{Zhao:2012p2116,  
Zhao:2012p2191}, one can make a change of phase space variables which makes 
the Hamiltonian and symplectic structures on the phase space well defined 
simultaneously at the price of introducing a non canonical symplectic structure. It is 
interesting to ask whether the corresponding statistical ensemble can be properly 
defined. Please notice that the systems studied in \cite{Shapere:2012p2114, 
Zhao:2012p2116, Zhao:2012p2191} belong exactly to the class of 
models for which the assumptions made in \cite{Ercolessi:2001p2304} are violated.
So, the study of statistical ensembles consisted of particles as described in 
\cite{Shapere:2012p2114, Zhao:2012p2116, Zhao:2012p2191} 
will enable us to see the consequence of violating the assumptions of 
\cite{Ercolessi:2001p2304}. Moreover, if the statistical ensemble for such 
strange mechanical systems turns out to possess some novel features
comparing to the standard canonical ensembles, it will provide for us an experimentally 
testable -- at least in principle -- effect for the spontaneous breaking of time translation
symmetry, about which tremendous interests is just emerging 
\cite{Chernodub:2012p1908, Gong:2012p2122, Shapere:2012p2115,
Shapere:2012p2280, Chernodub:2012p2335, Ghosh:2012p2173}.

\section{Strange description of Newtonian particles}

To facilitate the discussions to be made below, let us first briefly review the 
assumptions of \cite{Ercolessi:2001p2304} based on which the invariance of the 
partition function (\ref{part}) with respect to different choices of Hamiltonian structures
was proven. Consider a one dimensional Hamiltonian system 
$\mathcal{H}=\mathcal{H}(p,q)$ with the standard symplectic structure given in terms
of the canonical symplectic form $d\Omega=dq \wedge dp$. It is easy to see that a 
change of Hamiltonian 
\begin{align}
\mathcal{H}\rightarrow\mathcal{H}'=F(\mathcal{H}) \label{hamtr}
\end{align}
together with a change of symplectic structure 
\begin{align}
d\Omega\rightarrow d\Omega'
=F'(\mathcal{H}) dq\wedge dp \label{symtr}
\end{align}
would yield the same set of Hamiltonian equations of
motion. The claim of \cite{Ercolessi:2001p2304} is that, provided $ F'(\mathcal{H}) 
>0$ on the whole phase space, the above change of Hamiltonian structures will not 
change the partition function $\mathcal{Z}$, and hence all thermodynamic quantities 
will remain unchanged. As remarked in \cite{Ercolessi:2001p2304}, an important 
consequence of the condition $ F'(\mathcal{H}) >0$ is that the number of critical 
points for the Hamiltonian is kept unchanged. Actually, the condition 
$F'(\mathcal{H}) >0$ implies more than that. It guarantees that the critical points 
of the new Hamiltonian are the images of those of the original Hamiltonian.

Now consider the following Lagrangian system,
\begin{align}
\mathcal{L}=\beta_0\left(\frac{1}{12}m^2{\dot x}^4+mU{\dot x}^2-U^2\right), 
\label{lag}
\end{align}
where $\beta_0>0$ is a constant which has dimension $[E]^{-1}$, 
$U=U(x)$ is a smooth function of $x=x(t)$. This is the special case 
of $f=m^2/12$ of the $fgh$ model studied in \cite{Shapere:2012p2114}. 
Naively, the canonical momentum associated with $x$ can be introduced as
\begin{align}
p=\frac{\partial \mathcal{L}}{\partial \dot x}
=\beta_0\left(\frac{1}{3}m^2{\dot x}^3+2mU \dot x\right). \label{mom}
\end{align}
It is clear that in regions for which $U(x)$ becomes negative, $\dot x$ is 
multivalued in 
$p$, rendering the Hamiltonian 
\begin{align}
\mathcal{H}\equiv p\dot x -\mathcal{L} \label{ham}
\end{align}
also multivalued in $p$. Notice that for the Lagrangian (\ref{lag}), 
\begin{align}
\frac{\partial^2 \mathcal{L}}{\partial {\dot x}^2} = 
\beta_0 \left(m^2 {\dot x}^2 +2mU(x)\right), \label{hess}
\end{align}
the existence of regions for which $U(x)$ becomes negative implies that 
$\frac{\partial^2 \mathcal{L}}{\partial {\dot x}^2}$ has zeros. Such Lagrangian
systems are called {\it singular systems} in the literature.

Actually, singular Lagrangian systems can be subdivided into different types. Let 
$\mathcal{L}(q^i, \dot q^i)$ be a generic Lagrangian involving generalized 
coordinates $\{q^i, i=1,\cdots,n\}$ and generalized velocities 
$\{\dot q^i, i=1,\cdots, n\}$. 
Define $\mathrm{Hess}(\mathcal{L})=\mathrm{det}
\left(\frac{\partial^2\mathcal{L}}{\partial \dot q^i \partial \dot q^j}\right)$. 
If $\mathrm{Hess}(\mathcal{L})$ does not have any zeros on the space of states,
the system is called {\em regular}. If, instead, 
$\mathrm{Hess}(\mathcal{L})=0$ everywhere on the space of states, the system is 
called {\em globally singular}, or simply {\em singular}. If $\mathrm{Hess}
(\mathcal{L})=0$ only on some {\em proper subset} on the space of states, the 
system is called {\em locally singular}, or {\em strange}, and the solution to the 
condition $\mathrm{Hess}(\mathcal{L})=0$ will be referred to as the strange locus. 

According to the above definition, the Lagrangian (\ref{lag}) is strange provided 
$U\le 0$. Please note that the strange locus for case with $U=0$ is different from 
that for the cases with $U<0$: in the former case the strange locus is consisted
only of points with $\dot x =0$, while in the latter cases the strange locus 
corresponds to $\dot x =\pm \sqrt{-\frac{2U(x)}{m}}$, i.e. the particle can have 
nontrivial motion along the strange locus.  

Let us have a more careful look at the strange system (\ref{lag}) with $U<0$. 
For this system, although one can still define the canonical symplectic structure
$d\Omega=dx\wedge dp$, the Hamiltonian is not well defined in terms of the canonical
variables $(x,p)$. Consequently, the partition function $\mathcal{Z}$ associated 
with the above system seems to be not well defined.
However, as was shown in \cite{Zhao:2012p2116} and \cite{Zhao:2012p2191}, 
there exists a natural Hamiltonian description for the Lagrangian system 
(\ref{lag}). The clue is to change from the canonical phase space variables to the
set $(x,v)$ (here $v=\dot x$) of non canonical variables. In terms of these, the 
Hamiltonian
(\ref{ham}) is well defined,
\begin{align}
\mathcal{H}=\beta_0\left(\frac{1}{2}mv^2+U(x)\right)^2, \label{ham2}
\end{align}
and the associated Poisson structure reads\footnote{Notice that when 
$\frac{1}{2}mv^2+U(x)=0$, the Poisson structure becomes singular. 
However, 
such singularities will not affect the usefulness of the Poisson structure, because the 
time evolutions of $x$ and $v$ evaluated using the Poisson structure are never singular.
In particular, the evolution equations are satisfied even at the singularities 
of the Poisson brackets.}
\[
\{x,v\}=\frac{1}{2m\beta_0}\left(\frac{1}{\frac{1}{2}mv^2+U(x)}\right),\qquad 
\{x,x\}=\{v,v\}=0.
\]
Therefore, at least formally, we can make use of the definition (\ref{part}) to 
construct a partition function for the ensemble of $N$ particles governed by the 
strange Lagrangian (\ref{lag}). The required symplectic form $d\Omega$ is given by 
the inverse of the Poisson structure\footnote{The singularities in the Poisson 
structure corresponds 
to zeros of the integration measure $d\Omega$. In 
\cite{Ercolessi:2001p2304}, the integrations in the partition function is performed 
under the condition that the critical points of the Hamiltonian are excluded. 
In our case, the singularities correspond exactly to the critical points of the 
Hamiltonian (\ref{ham}). Since the integration measure becomes null at the critical 
points, their contribution to 
the integration is automatically excluded.}
\begin{align}
d\Omega = 2m\beta_0\left(\frac{1}{2}mv^2+U(x)\right) dx \wedge dv. 
\label{vol2}
\end{align}
Note that the unusual coefficient in front of $dx\wedge dv$ 
in (\ref{vol2}) can also be understood as 
the Jacobian arising from the change of integration variable $dp \rightarrow dv$.

Interestingly, the Hamiltonian equations of motion in this alternative description 
are identical to those of a canonical Newtonian particle, i.e.
\begin{align}
\dot x = v, \qquad m\dot v = - \frac{dU}{dx}. \label{equ}
\end{align}
So, the same set of equations of motion can also be obtained from the standard 
canonical Hamiltonian structure (here $\tilde p=mv$)
\begin{align}
&\mathcal{H}_1 =\frac{\tilde p^2}{2m} +U(x), \label{h1}\\
&\{x,\tilde p\}_1=1, \qquad \{x,x\}_1=\{\tilde p,\tilde p\}_1=0,\\
&d\Omega_1=dx\wedge d\tilde p. 
\end{align}

The two Hamiltonian descriptions $(\mathcal{H}_1, \{\cdot,\cdot\}_1)$ and 
$(\mathcal{H}, \{\cdot,\cdot\})$ for the same set (\ref{equ}) of equations of 
motion will henceforth be referred to as the first and the second 
Hamiltonian descriptions. 
Note that the two Hamiltonian descriptions also holds in the case $U=0$. What makes 
the $U=0$ case exceptional lies in that, for $U=0$, $\dot x$ is actually 
single-valued in $p$, so $\mathcal{H}$ is also single-valued and well 
defined in $(x,p)$. Consequently we can associate with the Hamiltonian
\begin{align*}
\mathcal{H}=p{\dot x}-\mathcal{L}
=\frac{1}{4}\beta_0{\dot x}^4=\frac{1}{4}\left(\frac{3p}{\beta_0 m^2}\right)^{4/3}
\end{align*}
the standard canonical Poisson bracket 
\begin{align}
\{x,p\}_1=1 \label{poisson3}
\end{align}
to make a Hamiltonian description $(\mathcal{H}, \{\cdot,\cdot\}_1)$ of the 
system. We have
\begin{align*}
&\dot x = \{x,\mathcal{H}\}_1=\left(\frac{3p}{\beta_0 m^2}\right)^{1/3},\\
&\dot p=\{p,\mathcal{H}\}_1=0.
\end{align*}
The first of these two equations gives the definition of $p$ in terms of $\dot x$, 
the second reproduces the constancy of $p$. This third Hamiltonian description 
for the $U=0$ case is actually not independent of the previous two descriptions. 
Writing $v=\dot x$, it can be inferred from (\ref{mom}) (by inserting $U=0$) 
that $p=\frac{1}{3}\beta_0 m^2 v^3$. Combining this and (\ref{poisson3}), we have
\begin{align}
\{x,v\}_1 = \frac{1}{\beta_0 m^2 v^2} =\{x,v\},
\end{align}
where $\{x,v\}$ is the Poisson bracket in the second Hamiltonian description. 
So the third Hamiltonian description for the case $U=0$ is actually identical to 
the second one and we are left with only two independent Hamiltonian descriptions
even for the $U=0$ case.

\section{Strange statistical ensemble}

In this section, we shall consider the statistical ensemble consisting of $N$ 
particles obeying the Newtonian equations of motion (\ref{equ}). 
For the ease of computation, let us consider the simplest choice $U(x)=-U_0 \le 0$. 
The choice $U\le 0$ is intentional, because only in 
this case the original Lagrangian (\ref{lag}) is strange. 

The partition functions under the first and second Hamiltonian descriptions
are given respectively as
\begin{align}
P_1=(\mathcal{Z}_1)^N,\quad \mathcal{Z}_1 &= \frac{1}{h} \int \mathrm{e}^{-
\beta \cdot\mathcal{H}_1} 
d\Omega_{1}, \label{z1}\\
P_2=(\mathcal{Z}_2)^N,\quad \mathcal{Z}_2 &= \frac{1}{h} \int \mathrm{e}^{-
\beta \cdot \mathcal{H}} 
d\Omega. \label{z2}
\end{align}
It is evident that the integrations in both (\ref{z1}) and (\ref{z2}) are well 
defined. What remains intact is whether these 
two integrations gives rise to the same result.

Before proceeding, let us remark that for constant potential $U(x)=-U_0$, although 
the two Hamiltonian descriptions appear to be very different, the associated 
Hamiltonian vector fields are actually the same. 
The canonical Hamiltonian vector field $\Gamma$ is given by
\[
\Gamma_1=\frac{\tilde p}{m} \frac{\partial}{\partial x},
\]
which, combined with the canonical symplectic form $d\Omega_1$, gives
\[
d \mathcal{H}_1  = \langle d\Omega_1, \Gamma_1\rangle 
= \frac{\tilde p}{m} d\tilde p.
\]
For the non canonical description, the Hamiltonian vector field is 
\[
\Gamma=v \frac{\partial}{\partial q},
\]
which yields
\[
d \mathcal{H}  = \langle d\Omega, \Gamma\rangle
= 2m\beta_0\left(\frac{1}{2}mv^2 + U\right) v dv.
\]
Considering the canonical Hamiltonian equations of motion, we can change 
$\tilde p$ into $mv$, then it is clear that 
\[
\Gamma_1=\Gamma,
\]
i.e. the very same $\Gamma_1$ is also the Hamiltonian vector field in the non 
canonical description. That the Hamiltonian vector fields remain the same in 
different Hamiltonian descriptions is a key ingredient \cite{Ercolessi:2001p2304}
in proving the invariance of partition functions under different Hamiltonian 
descriptions.

Since the integrands do not depend on $x$, the two integrations in  (\ref{z1}) 
and (\ref{z2}) can be easily evaluated. Assuming $x\in \left[-\frac{L}{2},\frac{L}
{2}\right]$, we have
\begin{align}
\mathcal{Z}_1&=\frac{L}{h}\left(\frac{2\pi m}{\beta}\right)^{1/2}
\mathrm{e}^{\beta U_0}, \quad U_0\ge 0.
\end{align}
Also, for  (\ref{z2}), we have 
\begin{align}
\mathcal{Z}_2|_{U_0>0}&=\frac{L}{h}\left(\frac{m}{2}\right)^{1/2}
\pi U_0^{3/2}\mathrm{e}^{-z}
\left(-I_{-1/4}(z)-I_{1/4}(z)+I_{3/4}(z)+I_{5/4}(z)
+\frac{1}{2z}I_{1/4}(z)\right), \label{z2s}
\end{align}
for $U_0>0$, where $z=\frac{1}{2}\left(\beta \beta_0\right) U_0^2$, $I_\alpha(z)$ is 
the well known modified Bessel function of the first kind,
which is always real valued for positive argument $z$. If $U_0=0$, the result is 
much simpler,
\begin{align}
\mathcal{Z}_2|_{U_0=0}&=(2m)^{1/2}\Gamma\left(\frac{3}{4}\right)
\left(\beta_0\beta\right)^{-3/4}.
\end{align}
It becomes clear that the partition functions in the two different Hamiltonian 
descriptions are very different.

\section{Why the difference occurs}

We have seen in the last section that the statement ``{\em the partition function is 
insensible to the form of the Hamiltonian as long as the equations of motion remain the 
same''}, which is quoted from \cite{Ercolessi:2001p2304}, becomes invalidated 
in our example system involving a strange Lagrangian description. 
It is natural to ask whether this is a consequence of violating the 
prerequisites of the arguments of \cite{Ercolessi:2001p2304} or if we had did something
wrong in the construction. 

Let us point out that the two descriptions presented in the last section 
{\it do} fit well in the form (\ref{hamtr}) and (\ref{symtr}). In deed, if we substitute
$\tilde p=mv$ in (\ref{h1}), then it follows immediately that
\begin{align}
\mathcal{H} = \beta_0(\mathcal{H}_1)^2, \label{sq}
\end{align}
and the symplectic volume element $d\Omega$ as given in (\ref{vol2}) is indeed
related to $d\Omega_1$ via (\ref{symtr}). What makes the relationship (\ref{sq})
differ from the framework of \cite{Ercolessi:2001p2304} is that the condition 
$F'(\mathcal{H}_1)>0$ is violated, because in this case, we have $F'(\mathcal{H}_1)
=2\beta_0\mathcal{H}_1$, whose range is $[-2\beta_0 U_0,\infty)$. Consequently, the 
number of critical points for the Hamiltonian is changed drastically before and 
after the change (\ref{sq}) of Hamiltonian descriptions, and most of the critical 
points of $\mathcal{H}$ are {\it not} the images of the critical point of 
$\mathcal{H}_1$. Actually, violation of the condition
$F'(\mathcal{H}_1)>0$ can happen not only to the quadratic map (\ref{sq}). Any  
choice of new Hamiltonian which is an even function of the original one will have 
the same effect, as long as the original Hamiltonian $\mathcal{H}_1$ can take 
negative values in some region in the phase space. Whenever $F'(\mathcal{H}_1)$ can 
have zeros, it would yield an algebraic equation for $\mathcal{H}_1$, whose solution
in terms of $v=\dot x$ will be nonzero in general. In such cases, the time 
translation symmetry in the ground state of $\mathcal{H}=F(\mathcal{H}_1)$ will be 
spontaneously broken. 

Careful readers might have already noticed something unusual when the partition 
function $\mathcal{Z}_2$ was evaluated. Indeed, putting in appropriate numeric 
values for each parameters, a negative value for  $\mathcal{Z}_2$ can arise from
(\ref{z2s}). Since the partition function is usually interpreted as the sum of 
(un-normalized) probability densities corresponding to each micro states in the 
phase space, a negative result seems difficult to understand. However, we argue 
that negativity of $\mathcal{Z}_2$ can be understood clearly if we take into account 
the fact that the momentum (\ref{mom}) is nonlinear in the generalized velocity $v$.
Inserting $U(x)=-U_0$ into (\ref{mom}), we can see that there are two local 
extrema of $p$  at $v=v_\pm =\pm\sqrt{2U_0/m}$. For $v\in (v_-,v_+)$, $p$ decreases 
as $v$ increases, therefore, $dp$ and $dv$ will have opposite 
signs in this region. Moreover, when $v\in (v_-, v_+)$, the value of the Hamiltonian
(\ref{ham}) is relatively small, yielding bigger Boltzmann weights 
$\mathrm{e}^{-\beta\cdot\mathcal{H}}$. When $v$ is beyond the above interval, the 
value of the Hamiltonian is relatively bigger, giving rise to smaller Boltzmann 
weights. Summing over all possible values of $v$, the partition function becomes 
negative. Let us stress that
the negativity of the partition function does not imply that the probability density 
becomes negative. It just reflects the fact that the symplectic volume element can 
become negative in the strange description. At this point, please also note that 
the path integral quantization for systems governed by a strange Lagrangian 
\cite{Anonymous:2012p1901} will also suffer from the same problem of negative 
measure. Nonetheless, the corresponding path integral is considered to be physically 
well established.

\section{Conclusions}

We have considered through a concrete example the construction of canonical 
ensemble consisting of $N$ particles admitting a strange Hamiltonian description. 
It turns out that the the partition function and hence the thermodynamic quantities 
in the strange description differ drastically
from their counterparts in the usual canonical Hamiltonian description. A violation of 
the condition $F'(\mathcal{H}_1)>0$ is identified as the reason for such differences.

Since there is an emerging interests in strange systems admitting a spontaneous 
breaking of time translation symmetry, it is important to ask what will be the 
experimentally falsifiable criterion for the occurrence of time translation symmetry 
breaking. In \cite{Shapere:2012p2114}, the perpetual motion in the ground state was
considered as the signature of the spontaneous breaking of time translation symmetry.
However, this statement depends on the choice of a specific Hamiltonian description,
because the ground states for the two Hamiltonians do not correspond to each other.
The perpetual moving states 
may be mapped to states with higher energies than the ground state by a change of 
Hamiltonian (e.g. mapping from $\mathcal{H}$ to $\mathcal{H}_1$ in the present 
context), thus eliminating the breaking of time translation symmetry. So, one cannot 
conclude that time translation symmetry is spontaneously broken by simply
observing states which undergo perpetual motion -- a physical mechanism which
determines the preferred choice of the Hamiltonian must also be involved. The 
research made in the present work suggests that one may look at the thermodynamic 
behavior
of the ensemble of strange systems. If one observes thermodynamic behaviors which
differ drastically from the predictions of standard canonical ensembles, 
then this observation might be used as a guidance for choosing a physical 
Hamiltonian description. In other words, the strange description and the standard
canonical description are inequivalent on the level of thermodynamics.

\section*{Acknowledgment} 

This work is supported by the National  Natural Science Foundation of 
China (NSFC) through grant No.10875059.  The author would like to thank 
X.-H. Meng for useful discussions.


\providecommand{\href}[2]{#2}\begingroup\raggedright\endgroup

\end{document}